\documentclass[fleqn,usenatbib]{mnras}

\usepackage{newtxtext,newtxmath}

\usepackage[T1]{fontenc}

\DeclareRobustCommand{\VAN}[3]{#2}
\let\VANthebibliography\thebibliography
\def\thebibliography{\DeclareRobustCommand{\VAN}[3]{##3}\VANthebibliography}

\pdfoutput=1 

\usepackage{graphicx}	
\usepackage{amsmath}	
\usepackage{upgreek}
\usepackage{caption}
\usepackage{subcaption}
\usepackage{verbatim}
\usepackage{newtxtext,newtxmath}
\newcommand{\dgr}{\,deg$^2$}
\newcommand{\trap}{\textsc{TraP}}
\newcommand{\pastro}{$p_{\textrm{astro}}$}
\newcommand{\mjy}{mJy\,beam$^{-1}$}
\newcommand{\ujy}{$\upmu$Jy\,beam$^{-1}$}

\title[LOFAR follow-up of O3 merger events]{LOFAR observations of gravitational wave merger events: O3 results and O4 strategy}
\author[K. Gourdji et al.]{
K. Gourdji,$^{1,2}$\thanks{E-mail: kgourdji@swin.edu.au}
A. Rowlinson,$^{3,4}$
R. A. M. J. Wijers,$^4$
J. W. Broderick,$^{5}$
A. Shulevski$^{3}$
\\
$^{1}$Centre for Astrophysics and Supercomputing, Swinburne University of Technology, Hawthorn VIC 3122, Australia\\
$^{2}$OzGrav: ARC Centre of Excellence for Gravitational Wave Discovery, Hawthorn VIC 3122, Australia\\
$^{3}$ASTRON, the Netherlands Institute for Radio Astronomy, Postbus 2, 7990 AA, Dwingeloo, The Netherlands\\
$^{4}$Anton Pannekoek Institute for Astronomy, University of Amsterdam, Science Park 904, 1098 XH Amsterdam, The Netherlands\\
$^{5}$International Centre for Radio Astronomy Research, Curtin University, GPO Box U1987, Bentley, WA 6845, Australia\\
}

\date{Accepted XXX. Received YYY; in original form ZZZ}

\pubyear{2021}

\begin{document}
\label{firstpage}
\pagerange{\pageref{firstpage}--\pageref{lastpage}}
\maketitle

\begin{abstract}
 The electromagnetic counterparts to gravitational wave (GW) merger events hold immense scientific value, but are difficult to detect due to the typically large localisation errors associated with GW events. The Low-Frequency Array (LOFAR) is an attractive GW follow-up instrument owing to its high sensitivity, large instantaneous field of view, and ability to automatically trigger on events to probe potential prompt emission within minutes. Here, we report on 144-MHz LOFAR radio observations of three GW merger events containing at least one neutron star that were detected during the third GW observing run. Specifically, we probe 9 and 16 per cent of the location probability density maps of S190426c and S200213t, respectively, and place limits at the location of an interesting optical transient (PS19hgw/AT2019wxt) found within the localisation map of S191213g. While these GW events are not particularly significant, we use multi-epoch LOFAR data to devise a sensitive wide-field GW follow-up strategy to be used in future GW observing runs. In particular, we improve on our previously published strategy by implementing direction dependent calibration and mosaicing, resulting in nearly an order of magnitude increase in sensitivity and more uniform coverage. We achieve a uniform $5\sigma$ sensitivity of $870$ \ujy\ across a single instantaneous LOFAR pointing's 21\dgr\ core, and a median  sensitivity of 1.1 \mjy\ when including the full 89\dgr\ hexagonal beam pattern. We also place the deepest transient surface density limits yet on of order month timescales for surveys between 60--340\,MHz (0.017\,deg$^{-2}$ above $2.0$\,\mjy\ and 0.073\,deg$^{-2}$ above $1.5$\,\mjy).
 
\end{abstract}




\section{Introduction}
\label{sec:intro}
Gravitational wave (GW) instruments have been detecting GW events from compact object mergers since 2015 \citep{ligo15}. The first GW merger event to consist of at least one neutron star, GW170817, was detected in the second observing run, in 2017 \citep{abbott17GW}. The event was temporally consistent with a short gamma-ray burst (GRB) detected by the \textit{Fermi} Gamma-ray Burst Monitor 1.7\, seconds later \citep{Goldstein17}. The binary neutron star merger's exceptionally small 30\,deg$^2$ localisation area facilitated swift identification of an optical counterpart \citep[][]{Coulter17}. This initiated a large-scale multiwavelength observing campaign of the electromagnetic (EM) counterpart that was eventually detected across the full EM spectrum \citep{abbott17EM}. This rich trove of data enabled several significant scientific advancements across a multitude of disciplines in astrophysics. Radio data in particular provided an unprecedented detailed view of a relativistic short gamma-ray burst jet and confirmed that GW170817 launched a successful jet that was viewed off-axis \citep[][]{Mooley18b}. 

In general, radio emission is predicted to arise at various stages of merging compact object systems that are composed of at least one neutron star (NS). Coherent emission has been predicted on timescales ranging from seconds before the merger to years after should the merger produce a stable NS remnant (see \citealp[][]{RowlinsonAnderson19} and \citealp[][]{Gourdji2020} for an overview of merger related models). Immediately following a NS merger, relativistic ejecta may be launched, which will shock the surrounding medium and power a synchrotron radio afterglow visible on timescales ranging from days to several months \citep[][]{PaczynskiRhoads93}. Such radio emission was observed for the first time from a GW event with binary NS merger GW170817 \citep[][]{Hallinan17}. Broadband radio monitoring of this afterglow was key to understanding the geometry, energy, kinematics and total energy of the relativistic outflow  \citep[e.g.][]{Hallinan17,Mooley18c,Mooley18b}. Notably, the geometry constraints broke degeneracies in GW distance measurements, leading to a significant improvement on the derived Hubble constant constraint \citep[][]{Hotokezaka19}. 


In addition to relativistic outflows, the merger will eject neutron-rich matter, which will also shock the surrounding medium to produce a long-lasting but fainter synchrotron afterglow, in addition to an optical kilonova \citep[][]{NakarPiran11}. The timescales of this afterglow will depend largely on the velocity distribution of the ejected matter. This ejecta is expected to have a distribution of mostly sub-relativistic velocities but with a relativistic fast tail that could contribute to the aforementioned early-time synchrotron emission \citep[][]{Kyutoku2014}. The majority slower moving ejecta will cause a radio afterglow that will peak at lower radio frequencies and at later times (year timescales) \citep[][]{Hotokezaka15,Hotokezaka16}. This type of afterglow has yet to be observed and searches for an associated re-brightening in GW170817 are ongoing \citep[e.g.][]{Balasubramanian22}.  



 The Low-Frequency Array (LOFAR) has been active in GW follow-up since the first GW observing run in 2015 (see \citealp{Gourdji22} for a summary). As we demonstrate in this paper, LOFAR's large instantaneous field of view and high sensitivity ($5\sigma$ of  870\,$\upmu$Jy across $\sim21$\,\dgr) make it an attractive tool for GW follow-up, where events containing a neutron star are localised to areas ranging from tens to, more typically, thousands of square degrees \citep{LVC_O3_1B,LVC_O3_2}. These large localisation areas challenge EM counterpart identification and indeed, thus far, GW170817 is the only event with a secure EM counterpart. Furthermore, the relatively low frequencies at which LOFAR observes for GW follow-up (between 115 and 189\,MHz, for single-beam pointings) can help to constrain the location of the synchrotron self-absorption frequency break in the highly time-dependent and degenerate afterglow spectrum \citep{Sari98}. Additionally, LOFAR's low-latency rapid response mode enables us to trigger on GW events and collect data within minutes, which probes potential prompt coherent emission \citep[e.g.][]{Rowlinson19, Rowlinson21}.
 

In this paper, we report on late-time LOFAR radio follow-up of three GW merger events containing at least one neutron star detected during the third GW observing. We use and build on the strategies developed in \citet{Gourdji22} to search for radio transients in wide-field LOFAR data corresponding to the radio afterglow of GW merger events. Here, we present a major improvement in reduction strategy relative to that presented in \citealp[][]{Gourdji22}: we perform direction dependent calibration and multi-beam mosaicing, resulting in a factor of 7 increase in median sensitivity across the beam pattern and more uniform coverage. While we do not find any significant radio transients, we present our strategy for the fourth observing run (O4) as well as the deepest wide-field slow transient survey at low frequencies. 

In Section \ref{methods}, we provide details about the merger events we observed, the LOFAR observations and data reduction. In Section \ref{results}, we present results from our transient search. We conclude and discuss our strategy for the upcoming GW observing run (O4), in Section \ref{discussion}.



\section{GW events, LOFAR observations and data reduction}
\label{methods}

The third GW observing run (O3) took place between 1 April 2019 and 27 March 2020, with a hiatus during the month of October 2019. The Kamioka Gravitational Wave Detector (KAGRA) joined the GW detector network on 25 February 2020. Following offline GW analysis of O3 data by the LIGO Scientific, Virgo and KAGRA (LVK) collaboration, there are 79 significant GW merger events \citep{LVC_O3_1B,LVC_O3_2}. These correspond to GW events with a false-alarm-rate less than 2 per year and probability of being astrophysical (\pastro) greater than 0.5. Of these significant events, 7 are consistent with containing at least one neutron star (1 BNS and 6 NS-black hole; NSBH\footnote{Two of the NSBH candidates could also be BBH systems with a low mass BH component.}). These are GW190425 (BNS), GW190814, GW190426\_152155, GW191219\_163120, GW200105\_162426, GW200115\_042309 and GW200210\_092254.  All events but GW191219\_163120 and GW200210\_092254 were detected during real-time analysis and were issued low-latency  public  alerts  via Gamma-ray Coordinate Network (GCN) Notices. No EM counterparts were robustly identified for any of the O3 candidates. 

We conducted late-time radio follow-up on a total of three merger events. One of these three GW events  is considered a significant GW merger candidate following offline analysis by LVK: NSBH candidate GW190426\_152155 (referred to as S190426c for the remainder of this paper). The events observed with LOFAR are S190426c (NSBH), S191213g (BNS), and S200213t (BNS). Note that the latter two events are not included in the final LVK catalogue, which was published approximately a year and half after their initial detections.

All LOFAR observations were conducted in the HBA Dual Inner configuration (24 core stations and 14 remote stations). 
Each multi-beam observation was centred at an observing frequency of 144.24\,MHz and has a total bandwidth of 13.48\,MHz. The two single-beam observations for S191213g were centred at 152.15\,MHz with a 72.22\,MHz bandwidth. Each subband contains 64 channels of width 3.051\,kHz. An integration time of 2 seconds was used for S190426c and S200213t, and 1 second for S191213g. Each observation was flanked by a ten minute calibrator scan of 3C48 and 3C196, in that order, except for the S200213t field, where 3C295 was used instead of 3C48. 

\begin{figure*}
     \centering
             \caption{Sky location probability density maps for the three GW events studied. \textbf{(a)} The corresponding LOFAR field ($4.6^{\circ} \times4.6^{\circ}$) is represented by a green box in the inset figure. The 50 and 90 per cent credible regions cover 235 and 1119 \dgr, respectively. There is a 9 per cent chance that the GW event is located within the LOFAR field. The localisation data come from GWTC-2.1 and correspond to products derived using the GSTLAL search pipeline.  \citep[][]{O3a_zenodo}. \textbf{(b)} The blue circle marks the location of the single LOFAR beam centered on the location of optical transient AT2019wxt. The localisation map is that derived by LALInference. \textbf{(c)} The full 89\dgr\ LOFAR mosaic is represented by the dashed blue line in the inset. The blue box in the inset corresponds to the $4.6^{\circ}\times4.6^{\circ}$ inner core of the LOFAR mosaic, where the sensitivity is uniform to within about 5 per cent. The 50 and 90 per cent credible regions cover 129 and 2325 \dgr, respectively. The core of the LOFAR mosaic captures 16 per cent of the probability density map. The localisation map is that derived by LALInference. These maps were created using \textsc{ligo.skymap}.}
        \label{fig:skymaps}
     \begin{subfigure}[t]{0.45\textwidth}
         \centering
        \includegraphics[width=\textwidth]{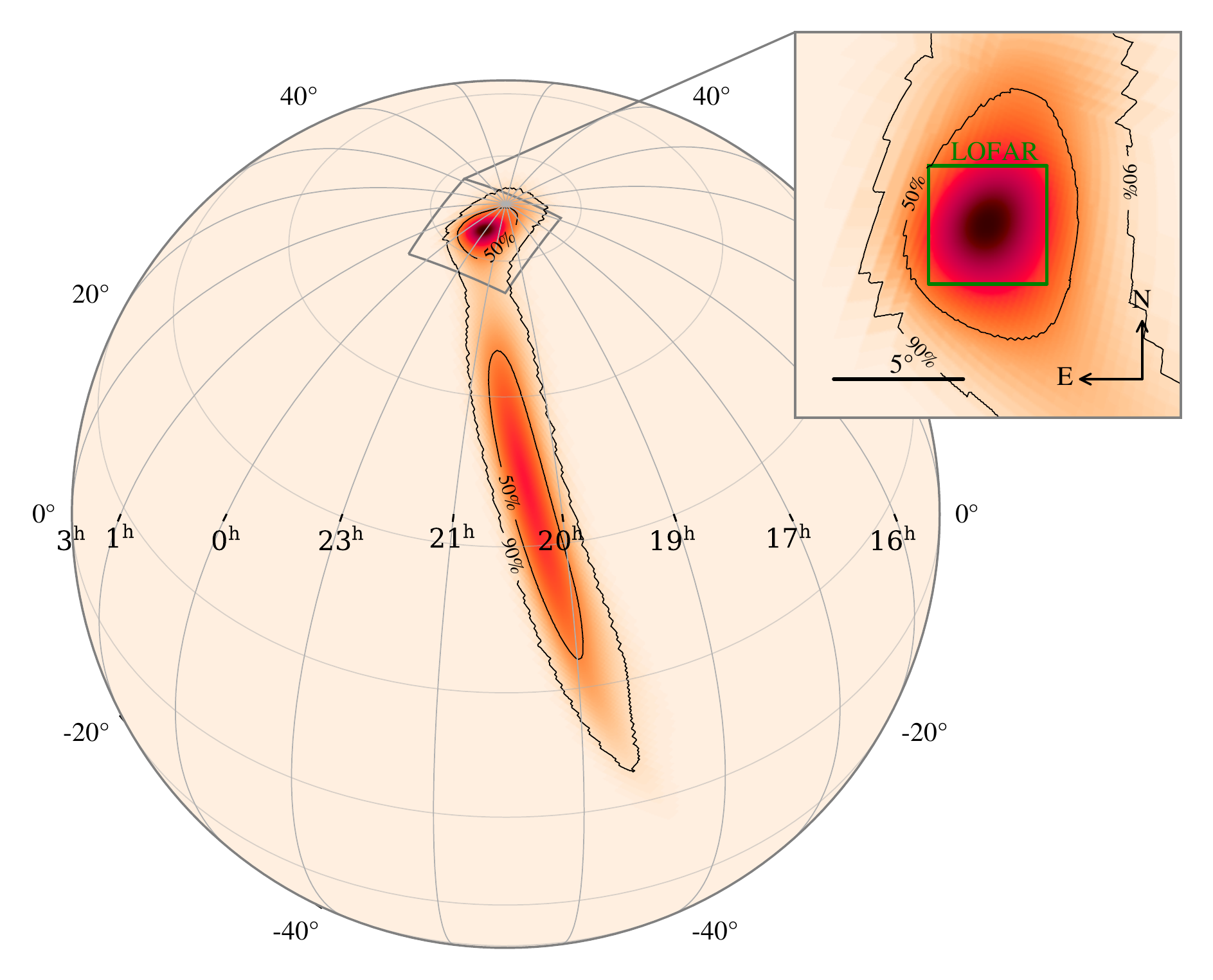}
        \caption{\textbf{S190426c}}
        \label{fig:S190426_map}
     \end{subfigure}
     \hfill
     \begin{subfigure}[t]{0.54\textwidth}
         \centering
         \includegraphics[width=\textwidth]{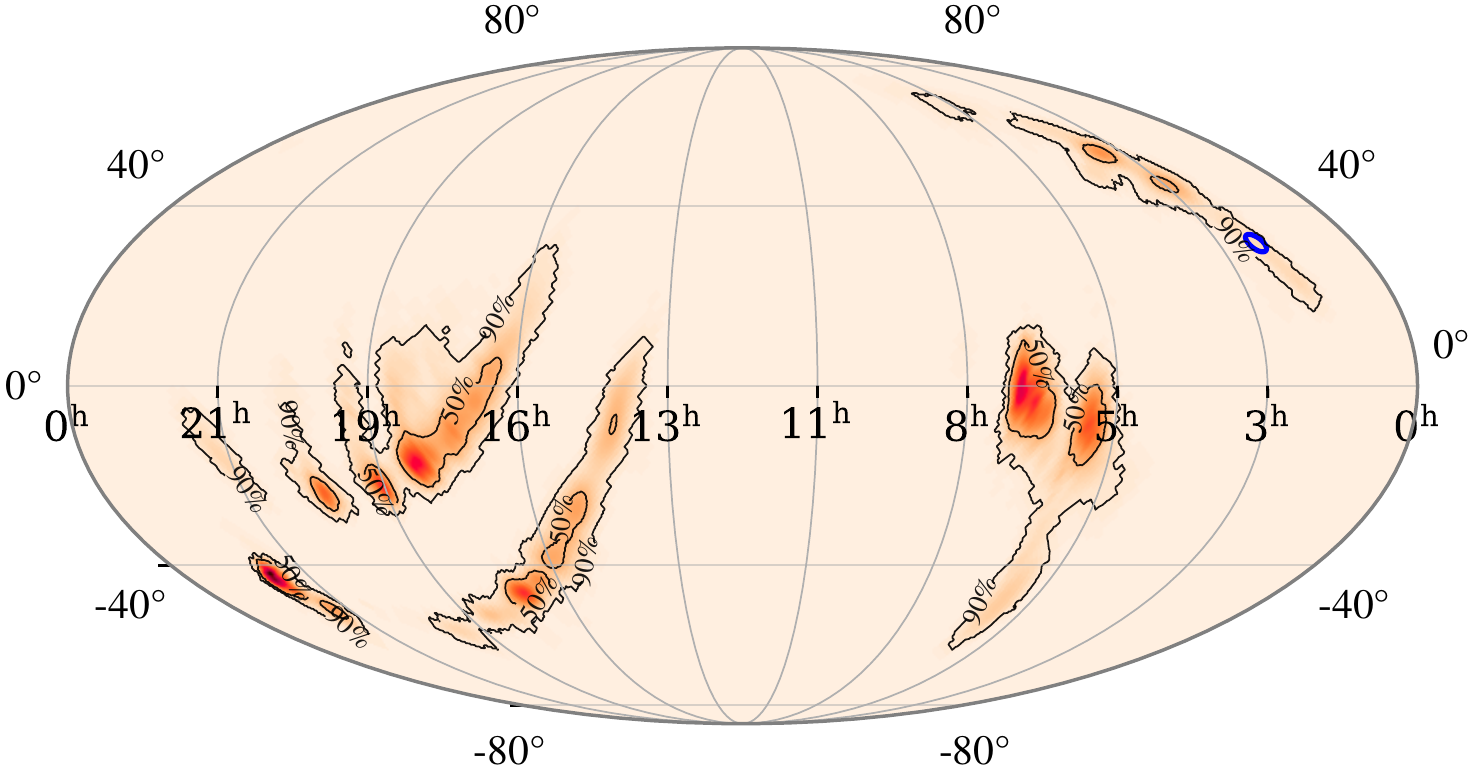}
         \caption{\textbf{S191213g}}
         \label{fig:AT2019wxt_map}
     \end{subfigure}
     \hfill
     \begin{subfigure}[b]{\textwidth}
         \centering
         \includegraphics[width=0.7\textwidth]{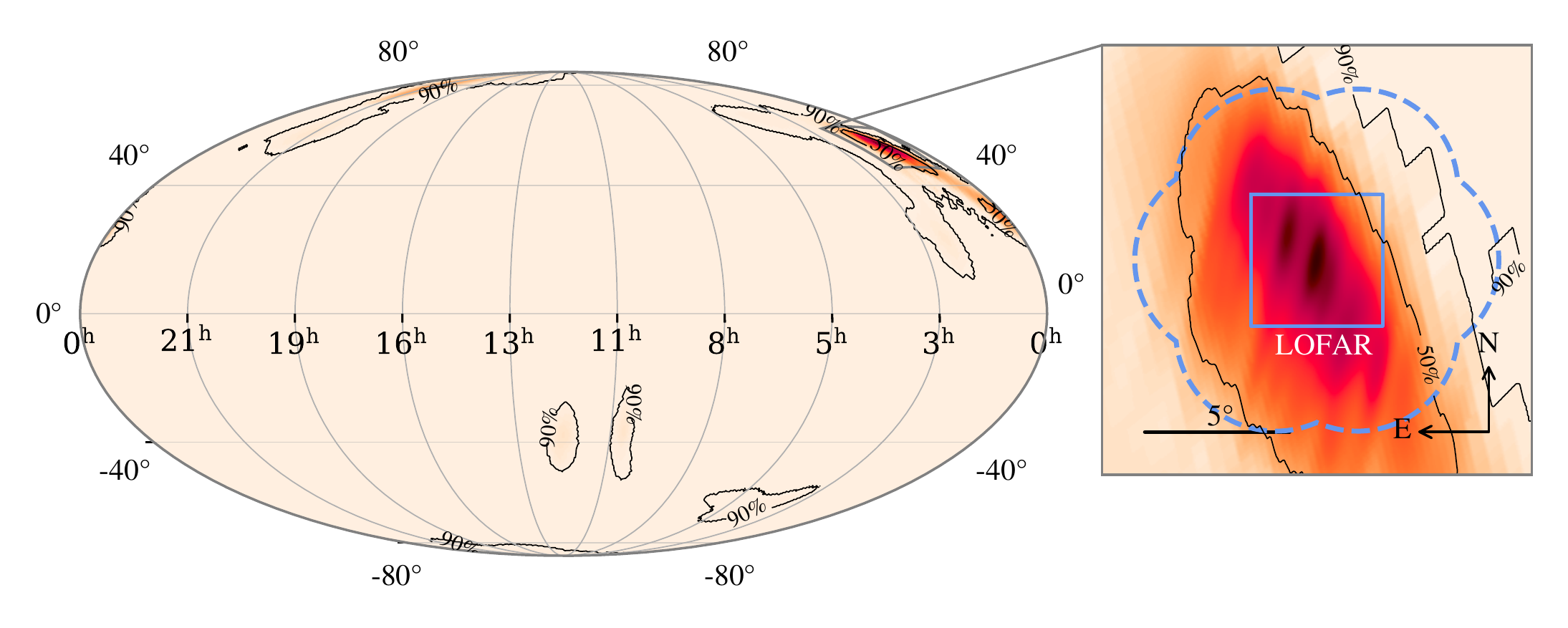}
         \caption{\textbf{S200213t}}
         \label{fig:S200213t_map}
     \end{subfigure}

\end{figure*}

\subsection{GW events and LOFAR observations}
\label{sec:obs}
\subsubsection{S190426c}
\label{obs:S190426c}
S190426c (a.k.a GW190426\_152155) is a NSBH star merger candidate that was detected by all three LIGO/Virgo detectors on 26 April 2019 at 15:21:55 UT \citep{S190426c_GCN}. A preliminary trigger alert was issued about 25 minutes later, stating that the event had a false alarm rate (FAR) of once per 1.63 years and was likely to be a merger system containing at least one NS and to have produced a merger remnant. This information, along with the event's location and the availability of an appropriate calibrator source, met the triggering criteria of our LOFAR rapid response strategy\footnote{We decreased the maximum allowed FAR to $10^{-14}$\,Hz in our triggering criteria partway through O3, to reduce the number of events that would trigger LOFAR. Hence the other two GW events did not trigger LOFAR rapid response observations.}. Unfortunately, a contemporaneous LOFAR observatory software update prevented triggered rapid-response observations of this event, which would have enabled us to be on-source within 5 minutes (see e.g. \citealp{Rowlinson19}). 

For late-time follow-up, we centered a single pointing comprising 7 overlapping LOFAR beams with centers separated by 2.8\textdegree, each with a radius of about 4.5\textdegree, on the peak of the location probability density map to produce a mosaic encapsulating 9 per cent of the location probability density map (see Figure \ref{fig:S190426_map}). Given that this field had not yet been observed by the LOFAR Two-metre Sky Survey \citep[LoTSS;][]{Shimwell17,Shimwell19}, we obtained 4 hours of interferometric LOFAR data one week post-merger on 3 May 2019, to create a reference image for transient searches in subsequent observations. We conducted a follow-up 4-hour observation approximately one month post-merger on 24 May 2019. The LVK later updated the classification of the event upon offline analysis and demoted the event from a 14 per cent chance of being terrestrial to a 58 per cent chance.  For this reason as well as the high FAR,  we cancelled further LOFAR follow-up of S190426c. The second release of the catalogue of merger events from the first half of O3 \citep{LVC_O3_1B} provides the latest GW analysis of this GW event. There, S190426c is considered to be a marginally significant candidate, with a FAR of $0.91$ per year and a probability of being astrophysical and a NSBH of 14 per cent according to the GstLAL pipeline\footnote{The low astrophysical NSBH probability is due to the low inferred rate of detectable NSBHs}.

\subsubsection{S191213g}
S191213g was detected by the LIGO--Virgo detectors on 13 December 2019 at 04:34:08 UT with a FAR of 1.1 per year and was deemed not to be a significant GW event in offline analysis \citep{LVC_O3_2}. S191213g was classified as a BNS merger candidate and the Pan-STARRS collaboration identified an interesting optical transient, PS19hgw/AT2019wxt, in its localization region on 16 December 2019 at 07:19:12 UTC \citep{AT2019wxt_GCN}. This led S191213g to be one of the O3 events with the most GCN Circulars, with follow-up across the EM spectrum and by neutrino facilities. We placed a single LOFAR beam at the location of AT2019wxt (RA = 01:55:41.941, DEC = +31:25:04.55, see Figure \ref{fig:AT2019wxt_map}) and obtained 8-hour observations 111 and 142 days post-merger. The field had already been processed by LoTSS and hence could be used as a reference. The optical transient was later classified as a likely type IIb supernova following its photometric evolution and spectroscopic studies, and a more recent multiwavelength analysis by \citet{Shivkumar22} classified AT2019wxt as an ultra-stripped supernova candidate.  


\subsubsection{S200213t}
S200213t was a low signal-to-noise event detected on 13 February 2020 at 04:10:40 UT by only the Hanford LIGO detector with a FAR of 0.56 per year. As with S191213g, S200213t was detected in real-time as a BNS but did not pass the LVK threshold to be considered a merger candidate in their offline analysis of GW data. We used a single LOFAR pointing comprising 7 overlapping beams (as described for S190426c above) centred on the peak of the location probability density map to probe 16 per cent of the map with the most sensitive part of the beam pattern (see Figure \ref{fig:S200213t_map}). We obtained 8-hour observations 6, 26 and 95 days post-merger. We ceased further follow-up given that the event did not meet the criteria of a GW merger candidate in offline analysis \citep[][]{LVC_O3_2}. 

\subsection{Data reduction}
Raw data were preprocessed by the observatory's data averaging pipeline following each observation.  Here, the visibilities were flagged for RFI using \textsc{aoflagger}, averaged in frequency and time, and bright nearby sources were subtracted where necessary using the pipeline's `demix' option. S190426c was averaged to 4 channels per subband and 4 seconds, and nearby Cas~A was demixed. S191213g was averaged to 4 channels per subband and 2 seconds. S200213t was averaged to 16 channels per subband and 2 seconds, to provide the flexibility to either demix or subtract Cas~A from the field during our offline processing.

The averaged data were then calibrated using \textsc{Prefactor} \citep[\texttt{v3.0};][]{prefactor}, which corrects for direction independent effects, and was used to demix Cas~A from our S200213t data. Calibration solutions from 3C196 were used for the first two fields, whereas 3C295 was used for S200213t. Skymodels from the TIFR GMRT Sky Survey (TGSS) were used for phase calibration of the target fields. The direction-independent calibrated visibilities were averaged to 2 channels per subband and subbands were concatenated into groups of ten and averaged in duration to 8 seconds. The \textsc{Prefactor} calibrated products were then passed through the \textsc{ddf-pipeline}\footnote{Second data release version. \url{https://github.com/mhardcastle/ddf-pipeline}} developed and used by the LoTSS team \citep{Shimwell19,Tasse21}, to deliver direction dependent calibration (DDC) and final image mosaics for scientific analysis\footnote{We used the \texttt{tier1-july2018.cfg} pipeline configuration file.}. The restoring beam of the final images has a radius of $3\arcsec$ and a pixel resolution of $1.5\arcsec$. The flux density scales of the images have been corrected in the way described in \cite{Shimwell19}. For S190426c, we analyse the inner 21\dgr core of the mosaic, where sensitivity is uniform to within 5 per cent. For S200213t, we analyze the full mosaic, with reduced sensitivity towards the outer edges, to demonstrate the strategy to be adopted for O4. Root mean square (RMS) noise values for each image are reported in Table \ref{tab:rms}.

\section{Transient search and results}
\label{results}
Image mosaics were run through the LOFAR transients pipeline \citep[\textsc{TraP}, version \texttt{r5.0};][]{swinbank2015} to extract sources and to identify new ones. Contrary to the default settings, we do not specify a source extraction margin, since the image quality of our mosaics is sound up to the image edges. Apart from that, default \trap\ settings were used, unless otherwise stated below. The method used for the transient search analysis is based on that developed in \citet{Gourdji22}. Due to improved image quality thanks to DDC, the transient candidate filters outlined there are no longer necessary and can be removed. For instance, we no longer reject candidates near to bright sources, given that DDC drastically diminishes the presence of (and hence false positives caused by) sidelobes. Furthermore, we can lower the source detection threshold from $7\sigma$ to $5\sigma$. The image sensitivities are summarised in Table \ref{tab:rms}.

\begin{table}
\begin{tabular}{lccccccc}
            & \multicolumn{2}{c}{S190426c}                  & \multicolumn{2}{c}{S191213g}                 & \multicolumn{3}{c}{S200213t}                                          \\ \hline \hline 
            Epoch & \multicolumn{1}{c}{1} & \multicolumn{1}{c}{2} & \multicolumn{1}{c}{1} & \multicolumn{1}{c}{2} & \multicolumn{1}{c}{1} & \multicolumn{1}{c}{2} & \multicolumn{1}{c}{3} \\ \hline
Mosaic core$^a$ & 231                    & 226                    &  --                    & --                    & 174                    & 153                    & 159                    \\
Full field$^b$  & --                      & --                      & --                      &    --                   &  216                     & 191                      &   190                    \\
AT2019wxt   &    --                   &    --                   &     63                  &  70                     &          --             &           --            &    --         \\ \hline 
\multicolumn{4}{l}{\small $^a$ Inner 21.15\dgr core of 7-beam mosaics.}\\
\multicolumn{4}{l}{\small $^b$ 89.11\dgr\ 7-beam mosaic.}\\
\end{tabular}

\caption{Image RMS values in units of $\upmu$Jy. Values are provided for each field's epoch image (numbered below the field names). For S190426c and S200213t, the median RMS values across the respective fields are reported. For S191213g, the RMS measured at the centre of the image, where AT2019wxt is located, is given.}
\label{tab:rms}
\end{table}

\begin{figure}
     \centering
     \begin{subfigure}[b]{0.5\textwidth}
         \centering
         \includegraphics[width=\textwidth]{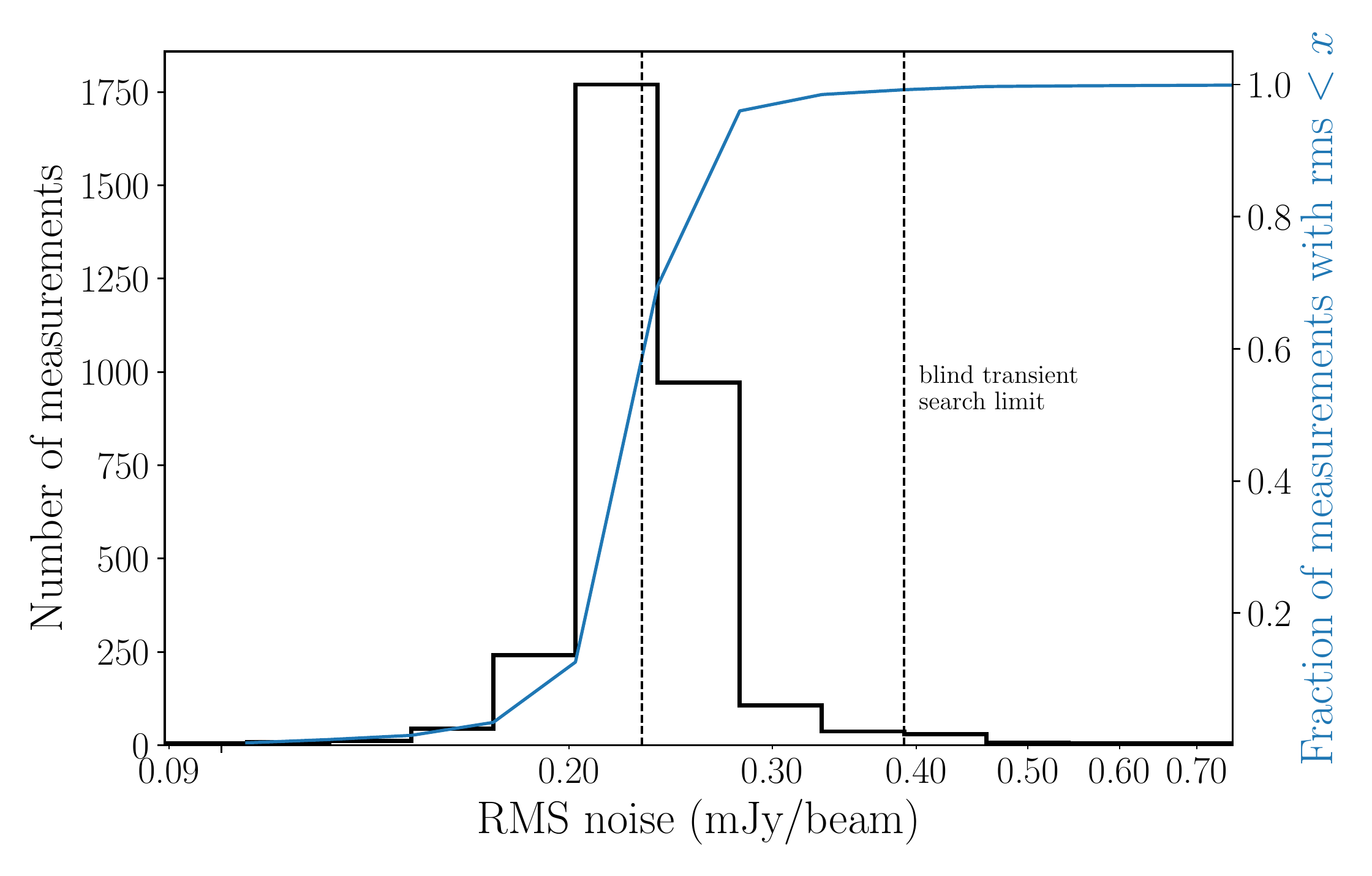}
     \end{subfigure}
     \hfill
     \begin{subfigure}[b]{0.5\textwidth}
         \centering
         \includegraphics[width=\textwidth]{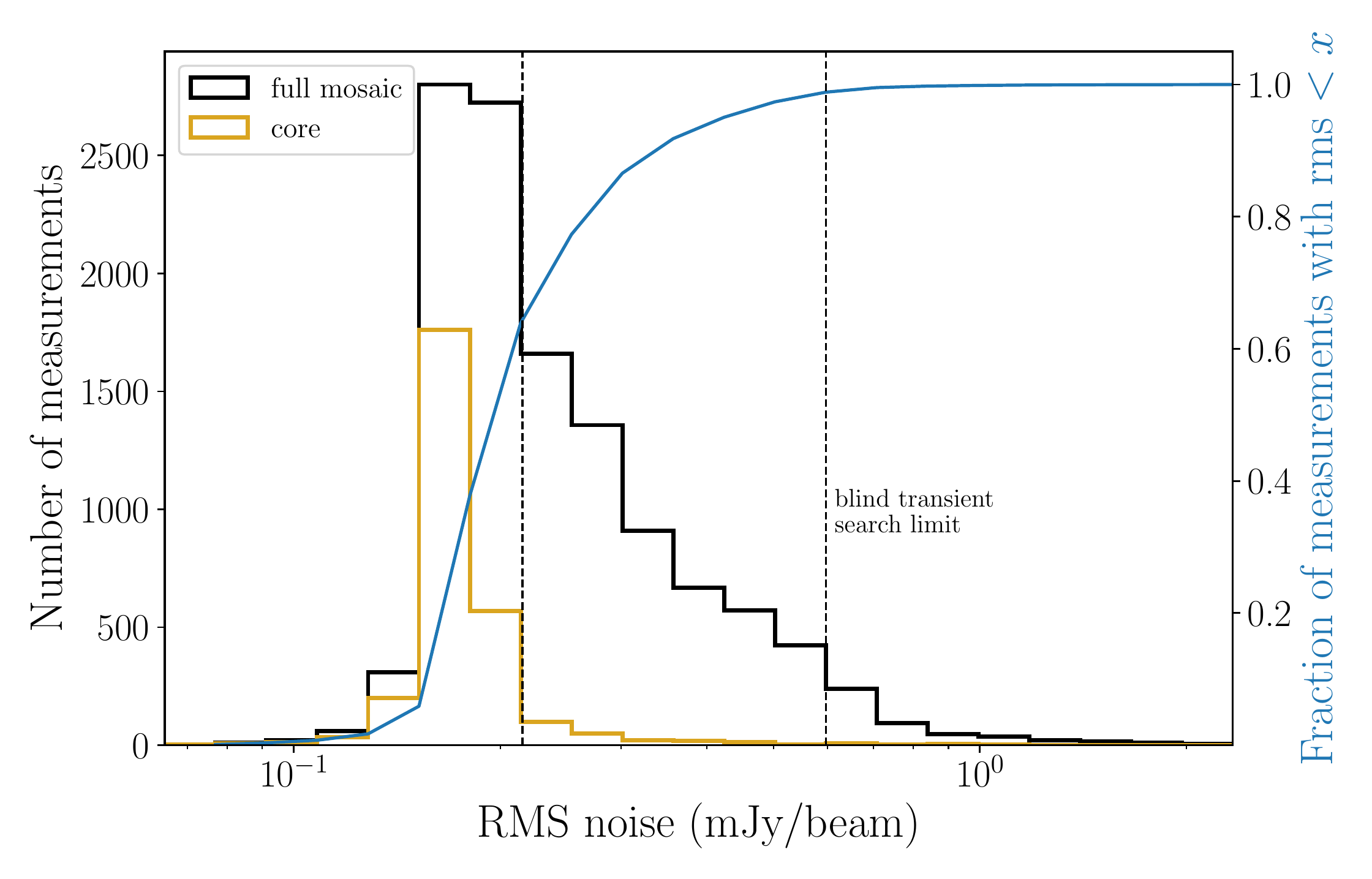}
     \end{subfigure}
     \caption{Distribution of RMS measurements from the sensitivity limiting images of S190426c (top) and S200213t (bottom). \textit{Top:} The median RMS is 231\,\ujy and is denoted by the left-most vertical line. Twenty one RMS measurements lie beyond the 700\,\ujy $x$-axis limit shown here, with a maximum value of 53\,\mjy. The right-most vertical line marks the RMS (0.39\,\mjy) below which 99 per cent of the measurements lie. This value is used to set the limiting sensitivity of 2.0\,\mjy\ for our blind transient surface density limit. The blue line traces the fraction of the image with RMS $<x$ (right-hand $y$-axis). \textit{Bottom:} Similar to the top panel. The distribution of RMS measurements in the inner 21\dgr core is included in yellow, while measurements from the full mosaic are shown in black. The left-most vertical line represents the median RMS of the full mosaic of 216\,\ujy. The 99 per cent blind transient limit is at 600\,\ujy. There are 23 additional RMS measurements beyond the 2.1\,\mjy\ limit applied to the plot for visualisation purposes.}
     \label{fig:rms}
\end{figure}

\subsection{S190426c}
\label{sec:190426c}
The average RMS in the first image of this field is slightly higher than in the second (see Table \ref{tab:rms}), and we thus consider the RMS properties of that image to determine the sensitivity of our search (Figure \ref{fig:rms}). Eighty per cent of the field has a measured RMS below 260\,$\upmu$Jy and the median $5\sigma$ sensitivity is 1.2\,\mjy.

\begin{figure}
    \centering

\begin{subfigure}[b]{0.47\textwidth}
\centering
    \includegraphics[width=\textwidth]{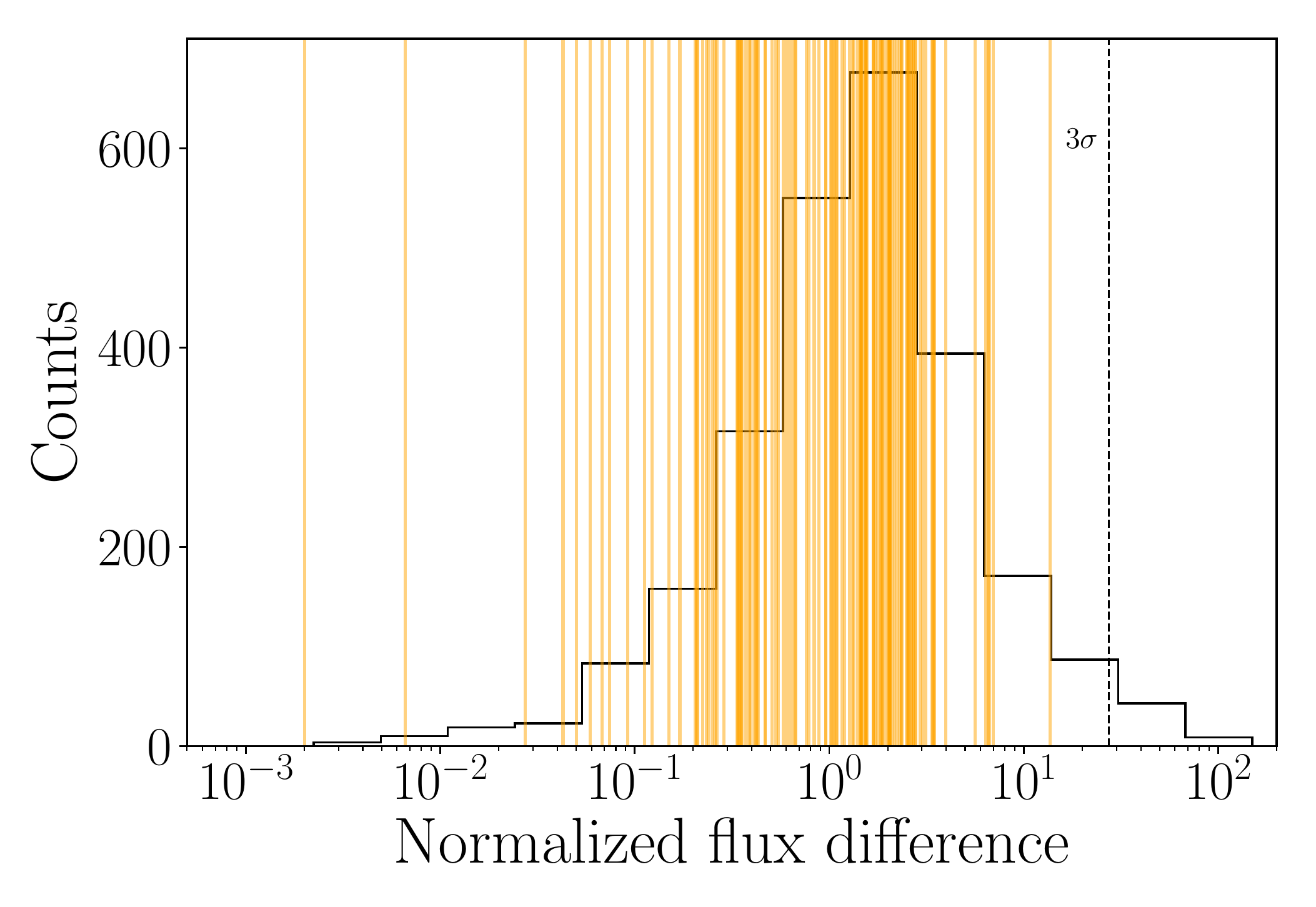}
\end{subfigure}
\hfill
\begin{subfigure}[b]{0.47\textwidth}
\centering
\includegraphics[width=\textwidth]{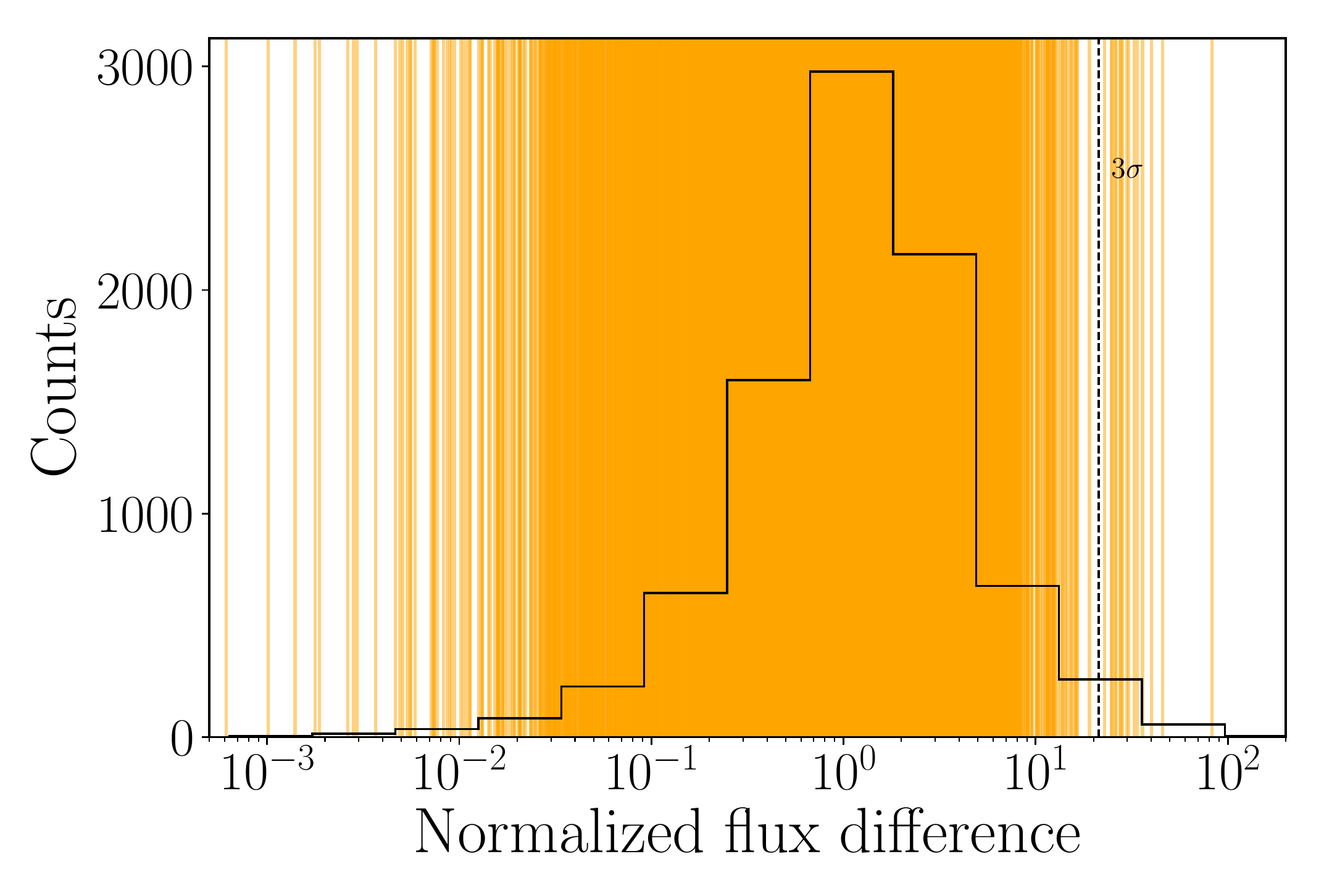}
\end{subfigure}
\caption{Normalized flux difference of transient candidates measured in each epoch image (orange) plotted over the distribution of normalized flux differences of persistent compact sources blindly extracted in the S190426c (top) and S200213t (bottom) fields. Thirteen transient candidates exceed three standard deviations from the persistent source distribution in the S200213t field and are further considered in the text.} 
\label{fig:fluxdiff}
\end{figure}

\trap\ extracted 4295 unique sources across the the two images,  of which 125 were automatically flagged as new sources. These correspond to sources that were blindly extracted in the second epoch which were not associated to any sources that were blindly extracted in the first epoch. The positions of these new sources were recorded into a monitoring list that was then passed onto \trap\ for a second run. This time, we forced \trap\ to extract sources at those positions in both images using the shape of the restoring beam. Following \citet{Gourdji22}, for each transient candidate position, we calculated the difference between the peak flux density measurement in each image normalized by their errors summed in quadrature. To check whether a flux difference is significant, we compared to the distribution of normalized flux differences calculated for all compact persistent sources (integrated to peak flux ratio less than 5) that were blindly extracted in the first \trap\ run. The results are shown in Figure \ref{fig:fluxdiff} where it is evident that none of the transient candidates are significant.  


\begin{figure*}
\centering
\includegraphics[width=0.75\textwidth]{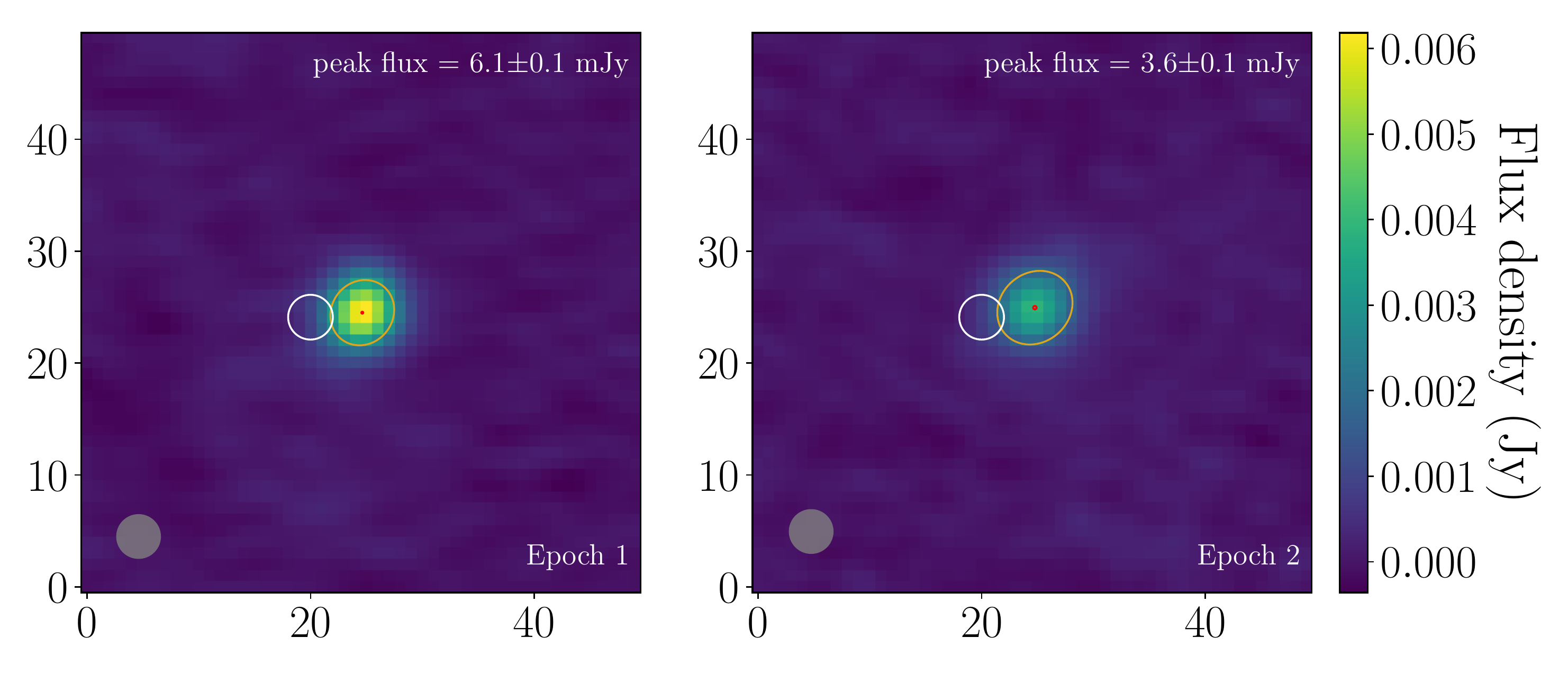}
\caption{LOFAR images of AT2019wxt and its host galaxy, KUG~0152+311, 108 and 139 days after the transient was identified (which itself was identified 3 days following GW event S191213g). AT2019wxt is denoted by a white circle with size corresponding to the synthesized beam. The host position and 2D Gaussian fit is denoted by the red and yellow markers, respectively, with values as measured by \trap\ during blind extraction. The circular synthesize beam is shown in the bottom left of each image.}
\label{fig:AT2019wxt_images}
\end{figure*}

\subsection{S191213g}
The RMS around the core of the first and second images, which is centred on AT2019wxt, is 63\,$\upmu$Jy and 70\,$\upmu$Jy, respectively. This sensitivity is consistent with the typical RMS achieved in LoTSS \citep{Shimwell19}. We performed a forced source extraction, taking the shape of the restoring beam, at the coordinates of AT2019wxt in our two images. The flux measured in the first and second epoch is $3.8\pm0.1$\,\mjy\ and $2.8\pm0.1$\,\mjy\, respectively. As can be seen in Figure \ref{fig:AT2019wxt_images}, the host galaxy of the transient is resolved and contaminates the extractions, which complicates accurately characterizing the flux density at the location of the transient. The resolved host galaxy is detected blindly by \trap\ in both images and the integrated flux is $12.6\pm0.3$\,\mjy\ in the first epoch and $10.0\pm0.4$\,\mjy\ in the second. The field was observed by LoTSS on 23 May 2017 and the host galaxy radio properties are recorded in the catalogue \citep{Shimwell22}. The resolved source has a peak flux of $5.7\pm0.1$\,\mjy, and integrated flux of $14.7\pm0.4$\,\mjy and a major axis of $10.2\pm0.2\arcsec$. 


We consider whether the apparent halving of the host galaxy's measured peak flux density in our images ($6.1\pm0.1$\,\mjy\ to $3.6\pm0.1$\,\mjy) is significant by comparing the flux difference normalized by the uncertainties (see \citealp{Gourdji22}) to the other persistent compact sources (with major axes less than 12$\arcsec$) blindly extracted in both of our images. We find that the normalized flux difference is consistent within a standard deviation of the distribution of persistent sources and hence conclude that the host's flux density variation is insignificant. We further note that the source extractor used by \trap, \textsc{PySE}, is not designed to handle extended sources \citep{pyse}. Thus, the errors on the integrated flux of the host reported by \trap\ are likely underestimated.

While this field turned out not to be actually related to a neutron star merger, this particular scenario, where the target of interest is contaminated by a nearby radio source, raises the question of whether the blind transient search strategy presented in \citet{Gourdji22} and used for the other fields in this study would succeed in identifying a merger radio counterpart that is blended with its host. Our transient search does not distinguish overlapping sources by, for example, fitting them with multiple overlapping Gaussians. Rather, as described in \citet{Gourdji22}, using its default settings, \trap\ would likely fit a single Gaussian to a multicomponent source, as it would to any extended source. A possible way to prevent this would be to use \trap's `deblend' option \citep{pyse}, though using this setting for such GW transient searches would require non-trivial testing and adjustments to our search algorithm and is beyond the scope of this study. 





\subsection{S200213t}
With its 8-hour integration time, this field represents the observing strategy to be adopted in O4 and we therefore search the full mosaic rather than the inner core to demonstrate the end-to-end strategy that will be used for events observed with LOFAR in the upcoming observing run. The transient search strategy used for this field is identical to that of S190426c, however the larger search area and additional epoch result in more transient candidates. The image from the first epoch has the highest RMS and so we use its properties to determine our transient search's median $5\upsigma$ sensitivity of 1.1\,\mjy\ ($870$\ujy in the mosaic core).

\trap\ blindly extracted 16257 unique sources across 3 images, each encompassing an area of 89\dgr. \trap\ identified 3002 transient candidates. These were treated following the same procedure as described in Section \ref{sec:190426c} and results from our candidate significance check is shown in Figure \ref{fig:fluxdiff}. Thirteen candidates have significant normalized flux differences and were inspected manually. All but one were clearly extended sources and visible in all three epochs and thus are rejected from further consideration. The remaining candidate is slightly resolved and detected in the second and third epochs ($25.0\sigma$ and $22.4\sigma$, respectively), but not present in the first (the forced fit, using the shape of the restoring beam, gives a peak flux of -$2.4\pm0.9$\,\mjy). However, large sidelobe image artefacts are present in the first image (as in the other two), which indicates that the source may have been erroneously cleaned during the direction dependent data reduction process. We confirm this by checking the intermediate calibration products and find that the source is present in the data up until the final \textsc{ddf-pipeline} calibration step. We thus rule out this candidate and conclude that no physical transients are present in this field. Diagnostic images of the transient candidate are included in the Appendix (Figure \ref{fig:trans_cand}). 

\subsection{Transient surface density}
\begin{figure}
\centering
\includegraphics[width=0.5\textwidth]{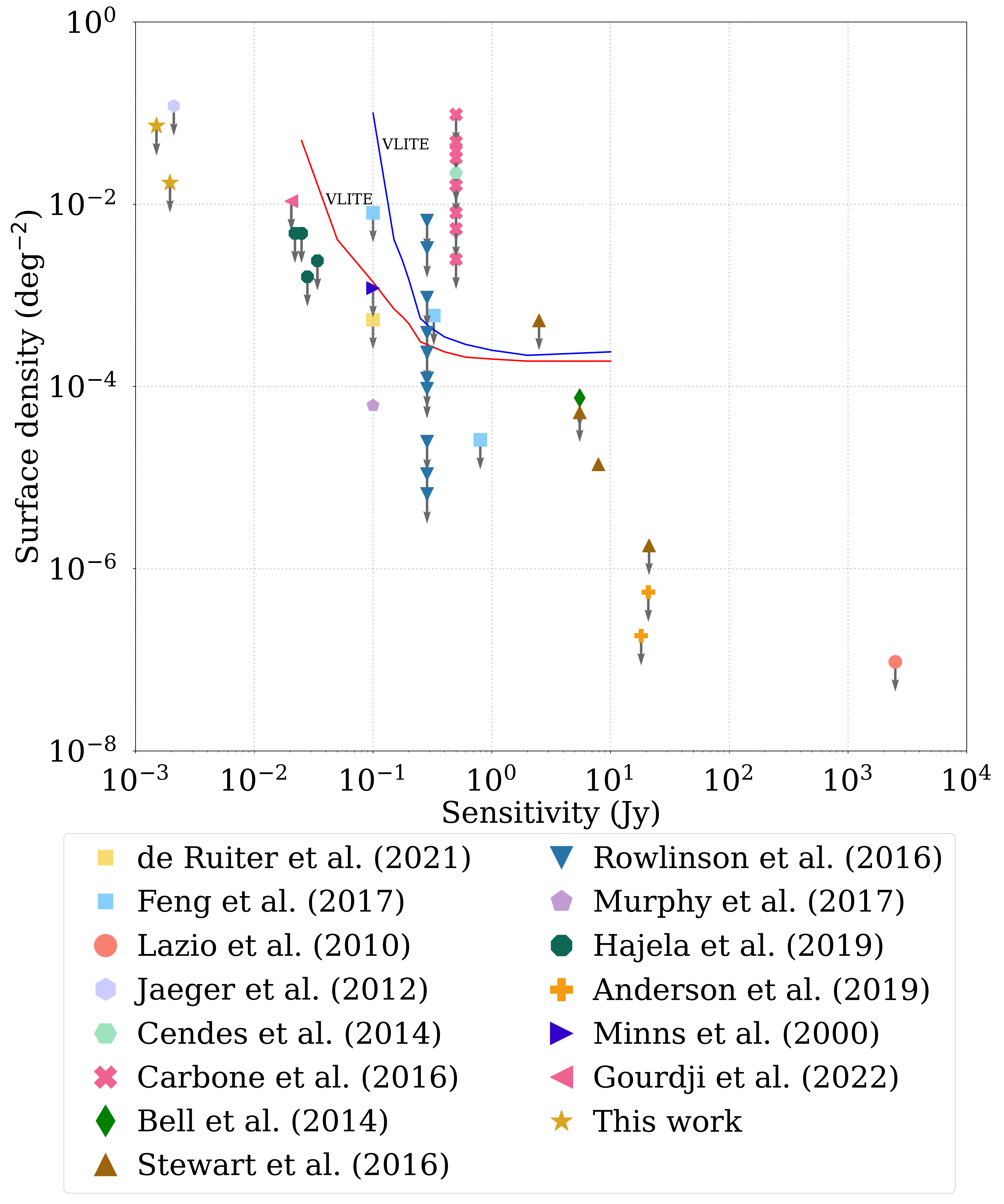}
\caption{Transient surface density limits for sources longer than one minute and surveys between 60\,MHz and 340\,MHz. Upper limits are represented by down-facing arrows. Limits from this study are denoted by gold stars and represent the most sensitive limits yet. The VLITE survey limits come from \citet{Polisensky2016}. This Figure was adapted from Figure 5 of \citet{Gourdji22}. Galactic centre radio transients are excluded.}
\label{fig:trans}
\end{figure}

To set a limit on the blind transient surface density, we consider a minimum sensitivity corresponding to 99 per cent completeness in each field considered. This corresponds to 2.0\,\mjy\ and 20.7\dgr\ of sky surveyed for S190426c, 3.0\,\mjy\ and 85.5\dgr\ for S200213t, and 1.5\,\mjy\ and 20.6\dgr\ for the latter's inner core. Following the same strategy as in \citet{Gourdji22}, we calculate the corresponding 95 per cent confidence surface density limits. Respectively, we find 0.15\,deg$^{-2}$, 0.018\,deg$^{-2}$ and 0.073\,deg$^{-2}$. It is, however, more constraining than the former two values to consider the surface density limit of the combined area surveyed above 2.0\,\mjy\ across both S190426c and S200213t ($20.7 + 76.4$ \,\dgr), which yields 0.017\,deg$^{-2}$. These constraints are summarised and compared to other relevant transient searches in Figure \ref{fig:trans}, where it is evident that this analysis constitutes the most sensitive blind search for transients on time scales longer than one minute at frequencies between 60\,MHz and 340\,MHz.

\section{Conclusions and O4 strategy}
\label{discussion}
We have presented a LOFAR search at 144\,MHz for radio transients related to three GW events detected in O3. We have demonstrated a significant improvement to the strategy previously presented in \citet{Gourdji22} by employing direction-dependent calibration and creating mosaics out of 7-beam pointings. We have achieved uniform 870\,\ujy\ $5\sigma$ sensitivity across the inner 21\dgr\ of the resulting 8-hour mosaics, and a median sensitivity of 1.1\,\mjy\ across the full 89\dgr\ mosaiced field. We have also demonstrated our ability to attain LoTSS sensitivities for single-beam fields (RMS $\sim70$\,\ujy). Further, the multi-epoch data-sets presented here constitute the deepest search for transients longer than one minute at frequencies between 60 and 340\,MHz.

The next GW observing run is expected to commence on 24 May 2023 and to last 18 months, with a likely month-long maintenance break part-way\footnote{\url{https://observing.docs.ligo.org/plan/index.html}}. This will be preceded by a month-long engineering run where exceptional events may be reported to the public. The next LOFAR observing cycle will be from 1 June 2023 to 31 May 2024, providing excellent overlap before stations are gradually brought offline thereafter as part of the LOFAR2.0 roll-out.

We plan to use the strategies used in the analyses presented here for LOFAR follow-up observations of events in O4. In particular, our wide-field search strategy will match that presented for S200213t ($870$\,\ujy sensitivity across an instantanenous 21\dgr\ field and further coverage at reduced sensitivity out to 89\dgr) and we will apply the S191213g strategy for events that have a previously identified EM counterpart. When available, LoTSS data will be used as reference for our transient search, and observations would be taken 1, 3, and 6 months post-merger. The cadence is chosen to properly sample the afterglow light curve and to probe events that are significantly off-axis (e.g. refer to Fig. 2 of \citealp{Broderick20}). Additionally, we plan to make use of LOFAR's rapid response capability to trigger on significant GW events, enabling us to gather interferometric data across 89\,\dgr\ within minutes of the merger. With these data-sets, we will use the strategy presented in \citet{Rowlinson19} and \citet{Rowlinson21} to search for coherent emission resulting from the merger remnant. The LVK forecasts $36^{+49}_{-22}$ BNS and $6^{+11}_{-5}$ NSBH merger alerts per year\footnote{\url{https://emfollow.docs.ligo.org/userguide/capabilities.html}} during O4, and we can expect about a quarter to have their location probabilities peak in LOFAR's field of view.


\section*{Acknowledgements}

We thank Robin Humble at OzSTAR HPC support for assisting in getting \textsc{ddf-pipeline} running on the OzSTAR computing cluster. We are grateful to Tim Shimwell for contributing to the data reduction of S190426c and for general support in using \textsc{ddf-pipeline}. We thank the observatory staff for coordinating LOFAR observations. KG acknowledges support through Australian Research Council Discovery Project DP200102243. AR acknowledges funding from the NWO Aspasia grant (number: 015.016.033)

This paper is based on data obtained with the International LOFAR Telescope (ILT) under project code LT10\_013. LOFAR \citep{vanhaarlem2013} is the Low Frequency Array designed and constructed by ASTRON. It has observing, data processing, and data storage facilities in several countries, that are owned by various parties (each with their own funding sources), and that are collectively operated by the ILT foundation under a joint scientific policy. The ILT resources have benefited from the following recent major funding sources: CNRS-INSU, Observatoire de Paris and Université d'Orléans, France; BMBF, MIWF-NRW, MPG, Germany; Science Foundation Ireland (SFI), Department of Business, Enterprise and Innovation (DBEI), Ireland; NWO, The Netherlands; The Science and Technology Facilities Council, UK; Ministry of Science and Higher Education, Poland.

\section*{Data Availability}

The LOFAR visibility data used in this analysis are publicly available at the LOFAR Long Term Archive (\url{https://lta.lofar.eu}) under project code LT10\_013. The image products used for the scientific analysis presented here will be shared on reasonable request to the corresponding author.


\nocite{deRuiter21,Murphy2017,Feng17,Stewart16,Carbone16,Bell14,Cendes14,Lazio10,Jaeger12,Hyman2009,MinnsRiley2000,Rowlinson16,Hajela19}

\bibliographystyle{mnras}
\bibliography{references} 



\clearpage
\captionsetup[subfigure]{labelformat=empty}
\captionsetup{labelfont=bf}
\appendix
\section{Transient candidate}
\noindent\begin{minipage}{\textwidth}
    \centering
         \centering
         \includegraphics[width=\textwidth]{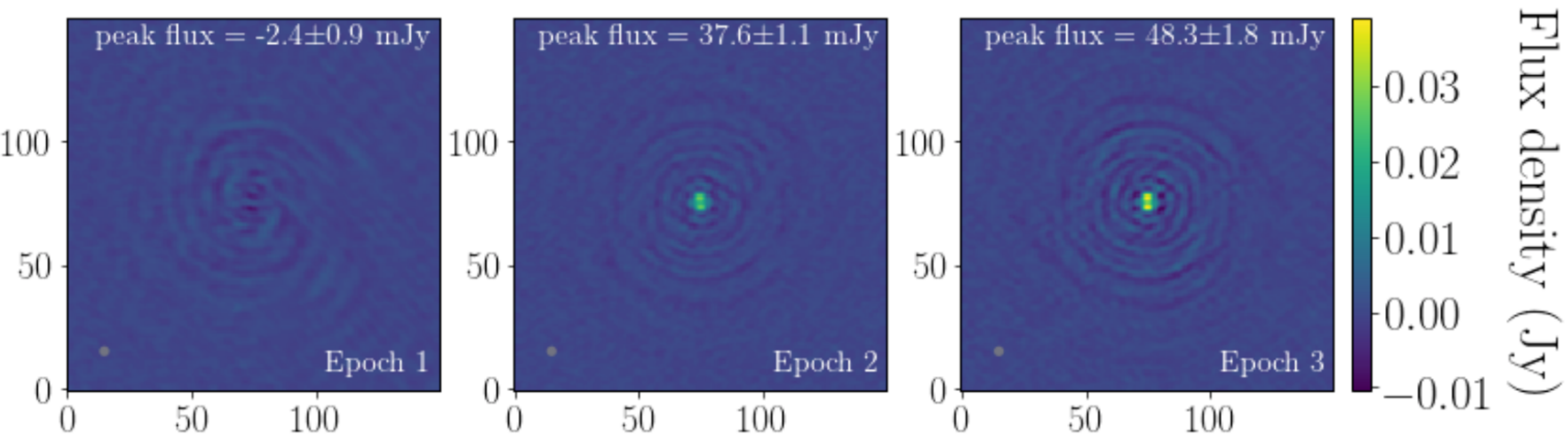}
     \bigskip
         \centering
         \includegraphics[width=0.5\textwidth]{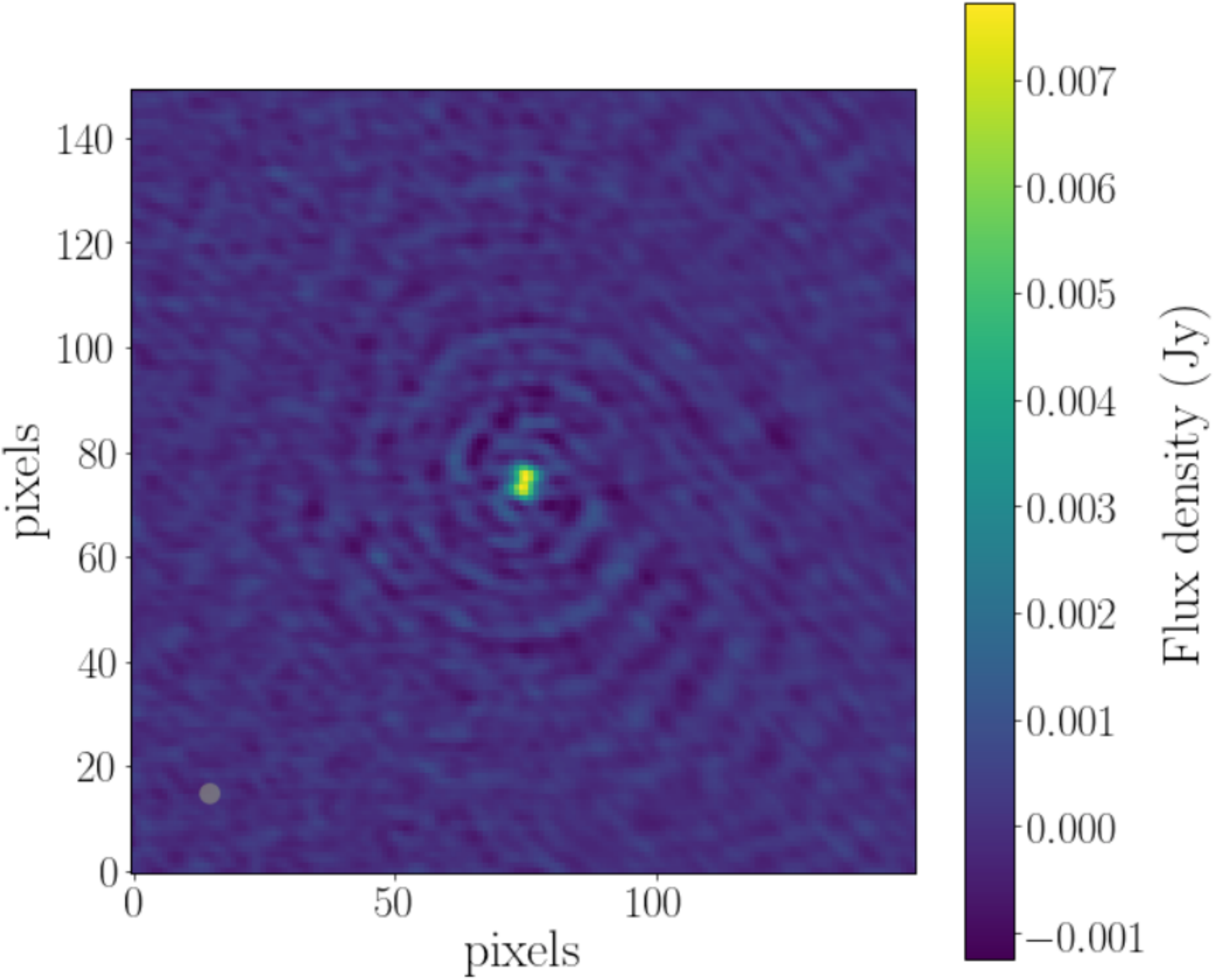}
     \captionof{figure}{Transient candidate detected in the S200213t field. \textit{Top:} Image cutouts at the location of the transient candidate in each epoch. The restoring beam is included in the bottom left corner of each panel. The source is detected blindly in the second and third epoch, and is not apparent in the first epoch. \textit{Bottom:} Cutout at the location of the transient candidate in an intermediate \textsc{ddf-pipeline} calibration step image of the first epoch. The source is clearly present and we thus conclude that the source was erroneously removed in the final calibration step and rule out this transient candidate.}
     \label{fig:trans_cand}
\end{minipage}


\bsp	
\label{lastpage}
\end{document}